# On-Policy Robust Adaptive Discrete-Time Regulator for Passive Unidirectional System using Stochastic Hill-climbing Algorithm and Associated Search Element

Mohsen Jafarzadeh[a], Nicholas Gans[b], Yonas Tadesse[a,c]

[a]Department of Electrical and Computer Engineering, The University of Texas at Dallas, Richardson, Texas 75080, USA

[b] UT Arlington Research Institute, The University of Texas at Arlington, Fort Worth, Texas 76118, USA

[c]Department of Mechanical Engineering, The University of Texas at Dallas, Richardson, Texas 75080, USA

**Abstract**

*Non-linear discrete-time state-feedback regulators are widely used in passive unidirectional systems. Offline system identification is required for tuning parameters of these regulators. However, offline system identification is challenging in some applications. Furthermore, the parameters of a system may be slowly changing over time, which makes the system identification less effective. Many adaptive regulators have been proposed to tune the parameters online when the offline information is neither accessible nor time-invariant. Stability and convergence of these adaptive regulators are challenging, especially in unidirectional systems. In this paper, a novel adaptive regulator is proposed for first-order unidirectional passive systems. In this method, an associated search element checks the eligibility of the update law. Then, a stochastic hill-climbing algorithm updates the parameters of the discrete-time state-feedback regulator. Simulation results demonstrate the effectiveness of the proposed method. The experiments on regulating of two passive systems show the ability of the method in regulating of passive unidirectional system in the presence of noise and disturbance.*

**Keywords**: adaptive control; on-policy optimization; digital control; discrete-time control; unidirectional system; passive system;

## 1. Introduction

Control of physical systems with a digital computer is common in practice [1-3]. Discrete-time state space representation of systems can be used to design digital controllers. A discrete-time state space model represents a system by one or more difference equation(s). In the case of observable, controllable, linear time-invariant (LTI) systems, a state-feedback controller can be used to improve the behavior of the system [4]. Many methods have been proposed to tune the parameters of a state-feedback controller, such as linear quadratic regulator (LQR) [5, 6], linear quadratic Gaussian (LQG) [7], dynamic programming [8-10], back-stepping [11], $\mathcal{H}_2$ [12], and $\mathcal{H}_\infty$ [13-17]. These methods require either offline system identification or a multi-physics model. In some cases, system identification process is not desired or possible. To address this issue, adaptive controllers have been used widely [18-21]. Adaptive controllers can improve performance in the case that the system dynamics change over the time.

Unidirectional systems are a special class of systems that feature inputs act only in one direction. A system can be unidirectional by inheritance or by design. For example, the force of an artificial muscle [22] is inherently unidirectional, and the speed of any permanent magnet direct current motor with a half-bridge driver is unidirectional by design. Many researchers use non-adaptive controllers for these type of system [23-28]. Unidirectional systems can be modeled as non-linear or switched linear by state space systems. Therefore, linear adaptive controllers may not work. Also, the non-linearity can cause instability in the closed-loop systems. The stability issue may be addressed by passivity concept [29]. A state space system is called passive if it is dissipative with respect to the supply rate. By using Lyapunov stability theory, a passive system is a stable system if the input is constant [30-33]. Also, if the additive energy of the controller is less than the dissipated energy of the passive plant, the closed-loop system is passive [34]. There has been several works presented on the control of artificial muscles that address different aspects depending on the actuation technologies [23, 35-40]. In this paper, we use the passivity property to design a stable controller.

Controllers can be divided in two categories: regulator and tracker. In a regulator controller, the reference trajectory is constant, $r(k) = r(k+1)$. Also, regulators can follow a trajectory when $r(k) \approx r(k+1)$. In the tracking controller, the current reference is independent from reference in next time step, i.e, $r(k) \perp r(k+1)$. In this paper, we only focus on regulators. We postpone tracking controller for future works.

The main contribution of this paper is a new method that uses an associated search element (ASE) for the stochastic hill-climbing algorithm to tune the parameters of a discrete-time state-feedback regulator for first-order passive unidirectional systems. As shown in Fig. 1, the stochastic hill-climbing algorithm updates the parameters of the state-feedback regulator. The associated search element is used to check the eligibility of adaptation as well as the step size and direction of the hill-climbing algorithm, such that it guarantees the stability of the system. The proposed method works for any first-order passive unidirectional system, which is confirmed experimentally and theoretically. It also experimentally verified to work on some higher order system. The proposed method is regulator, $r(k+1) = r(k)$, and it is not tracker.

This paper is organized as follows. In Section 2, the discrete-time state space model of a general first order passive unidirectional system is stated. The next Section describes related works. In Section 4, the proposed method is described in details. Stability proof of the proposed method is shown in the Section 5. Section 6 demonstrates the simulation results of the proposed method as compared with other methods. Experiment results on regulating of the force of a TCP muscle and an SMA muscle as examples of passive unidirectional systems are illustrated in Section 7. The last section is a conclusion for adaptive regulator of passive unidirectional systems by the stochastic hill-climbing algorithm and associated search element.

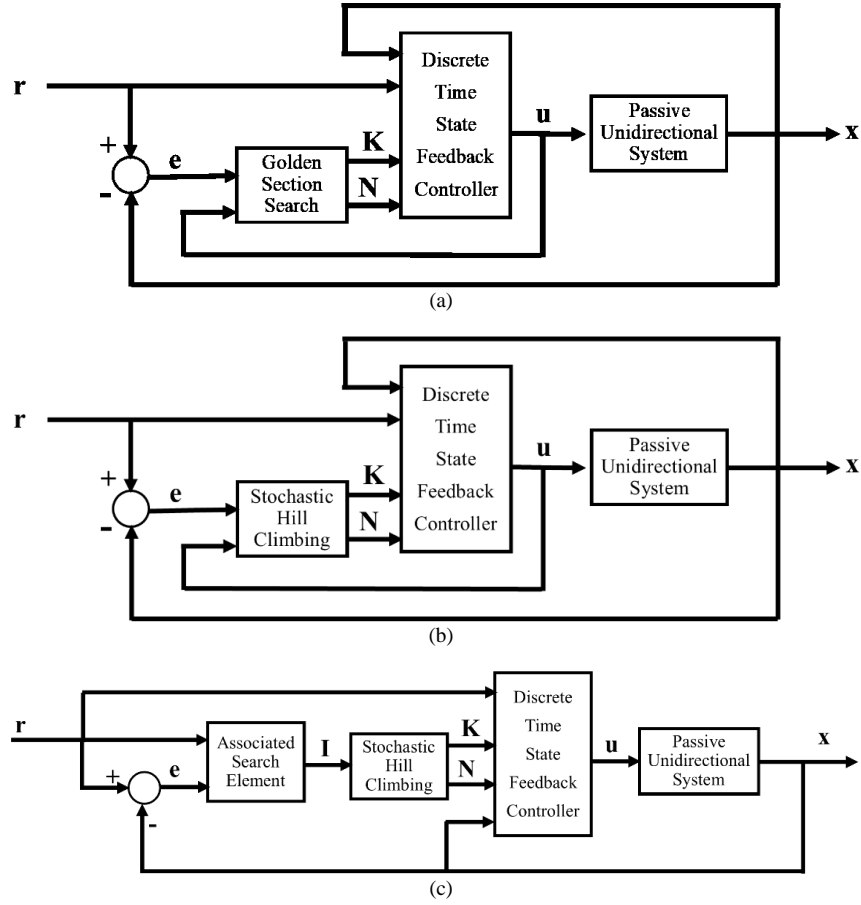

Fig. 1. Block diagram of an adaptive discrete-time state-feedback regulating system, $r(k+1) = r(k)$, with (a) golden section search (b) stochastic hill climbing (c) proposed method, stochastic hill-climbing with associated search element.

## 2. State-feedback regulator

A first-order unidirectional system can be modeled by a discrete-time state space equation as

$$x(k+1) = \begin{cases} a\,x(k) + b\,u(k) &, u(k) > 0 \\ f\,x(k) &, u(k) = 0 \end{cases} \qquad (1)$$
$$y(x) = c\,x(k)$$

where, $x(k) \in \mathbb{R}$ is the state of the system at time t= $kT$. The $T \in \mathbb{R}_{>0}$ is the sampling time. The $u(k) \in \mathbb{R}_{\geq 0}$ is the input of the system. The $a \in \mathbb{R}$, $b \in \mathbb{R}_{>0}$, $c \in \mathbb{R}$ and $f \in \mathbb{R}$ are scalars. The proposed method is general, in that it works for any first-order passive unidirectional system. The input and output can be anything. For example, for a DC motor, we will measure the velocity. For a light bulb, we will measure the light intensity. For TCP, we will measured the force. From passivity properties, we know that absolute values of $a$ and $f$ are less than one. The regulating error $e(k)$ is defined as the difference between the system state and desired state $r(k)$

$$e(k) = r(k) - x(k). \qquad (2)$$

Therefore,

$$e(k+1) = r(k+1) - x(k+1). \qquad (3)$$

By substituting Eq. (1) in Eq. (3)

$$e(k+1) = \begin{cases} r(k+1) - a\,x(k) - b\,u(k) &, u(k) > 0 \\ r(k+1) - f\,x(k) &, u(k) = 0 \end{cases}. \tag{4}$$

From Eq. (2), we can write

$$\begin{cases} -a\,e(k) = -a\,r(k) + a\,x(k) \\ -f\,e(k) = -f\,r(k) + f\,x(k) \end{cases}. \tag{5}$$

By adding Eq. (4) and Eq. (5)

$$\begin{cases} e(k+1) - a\,e(k) = r(k+1) - a\,r(k) - b\,u(k) &, u(k) > 0 \\ e(k+1) - f\,e(k) = r(k+1) - fr(k) &, u(k) = 0 \end{cases}. \tag{6}$$

Thus,

$$e(k+1) = \begin{cases} a\,e(k) + r(k+1) - a\,r(k) - b\,u(k) &, u(k) > 0 \\ f\,e(k) + r(k+1) - fr(k) &, u(k) = 0 \end{cases}. \tag{7}$$

The state-feedback regulating law is defined as

$$u(k) = \begin{cases} -K\,x(k) + N\,r(k) &, e(k) > 0 \\ 0 &, e(k) \leq 0 \end{cases} \tag{8}$$

where $K \in \mathbb{R}_{\geq 0}$ and $N \in \mathbb{R}_{\geq 0}$ are positive scalars. We know that $N > K$ because we used passive regulator. From (8), we can conclude

$$\begin{cases} u = 0 \Rightarrow e(k) \leq 0 \\ u > 0 \Rightarrow e(k) > 0 \end{cases} \tag{9}$$

and rewrite Eq. (9) as

$$e(k+1) = \begin{cases} a\,e(k) + r(k+1) - a\,r(k) - b\,u(k) &, e(k) > 0 \\ f\,e(k) + r(k+1) - fr(k) &, e(k) \leq 0 \end{cases}. \tag{10}$$

By substituting Eq. (2) in Eq. (8)

$$u(k) = \begin{cases} K\,e(k) + (N-K)\,r(k) &, e(k) > 0 \\ 0 &, e(k) \leq 0 \end{cases}. \tag{11}$$

By substituting Eq. (11) in Eq. (10)

$$e(k+1) = \begin{cases} [a - bK]\,e(k) + r(k+1) + [bK - bN - a]\,r(k) &, e(k) > 0 \\ f\,e(k) + r(k+1) - fr(k) &, e(k) \leq 0 \end{cases}. \tag{12}$$

In any regulating system we have $r(k+1) = r(k)$, therefore

$$e(k+1) = \begin{cases} [a - bK]\,e(k) + [1 + bK - bN - a]\,r(k) &, e(k) > 0 \\ f\,e(k) + [1 - f]\,r(k) &, e(k) \leq 0 \end{cases}. \tag{13}$$

Thus, to minimize the absolute value of the error, the following equation must be satisfied

$$N = K + \frac{1-a}{b} = K + K_{DC}. \tag{14}$$

Then,

$$e(k+1) = \begin{cases} [a - bK]\,e(k) &, e(k) > 0 \\ f\,e(k) + [1 - f]\,r(k) &, e(k) \leq 0 \end{cases}. \tag{15}$$

The parameter $K$ affects the transient state of the closed-loop system response. The $N - K$ affects the steady state response. If we know the exact values of $a$ and $b$, we can optimize $K$ and $N$ by linear quadratic regulator [5]. If we do system identification for $a$ and $b$ and estimate them by $\hat{a}$ and $\hat{b}$, we can optimize $K$ and $N$ by $H_\infty$ [13, 14]. However, in some application, offline estimations of $\hat{a}$ and $\hat{b}$ are neither possible or nor economical. Also, in some application, $a$ and $b$ are slowly varying with time such that after some time, the initial estimation is no longer valid. Therefore, in this paper, we proposed how to optimize the $K$ and $N$ during operation when the parameters $a$ and $b$ are unknown.

Because the proposed method is regulator, $r(k+1) = r(k)$, we will optimize the Eq. (15) when Eq. (14) holds. To design a tracker, we will find an optimization algorithm for equation Eq. (15) in future work.

## 3. Related works in optimization of regulators

The goal of the regulators is to minimize regulating error Eq. (15). The cost of the system is defined as

$$J = \sum_{k=1}^{T} |e(k)| \tag{16}$$

where $T$ is the number of step in each episode. There is always a physical bound in the magnitude of the regulators' output. Therefore, the following optimization problem can be used to tune the regulator parameters.

$$\begin{bmatrix} N^* \\ K^* \end{bmatrix} = \begin{cases} \underset{N,K}{\mathrm{argmin}} \sum_{k=1}^{T} |e(k)| \\ s.t. \underset{0<k<T-1}{\max} u(k) < \mathcal{U} \end{cases} \tag{17}$$

where $\mathcal{U}$ is the maximum limit of the regulator output. The $N^*$ and $K^*$ are optimal (local or global) values of $N$ and $K$. When a method cannot handle the above constraint, the following cost function can be used.

$$J' = \sum_{k=1}^{T}[|e(k)| + \alpha\, u(k)] \tag{18}$$

where $\alpha \in \mathbb{R}_{>0}$ is a scalar. Therefore,

$$\begin{bmatrix} N^* \\ K^* \end{bmatrix} = \underset{N,K}{\mathrm{argmin}} \sum_{k=1}^{T}[|e(k)| + \alpha\, u(k)]. \tag{19}$$

In mathematics and computing science, many algorithms exist to find the roots of an equation numerically, such as bisection method, regula falsi method, Secant method [41], Ridders' Method, Van Wijngaarden–Dekker–Brent Method, Muller's method, Broyden's method, and inverse quadratic interpolation method [42]. To use these methods, having the partial derivative of the cost function with respect to N and K is necessary. However, the cost functions Eqs. (16) and (18) are not analytical functions.

Kiefer proposed the Fibonacci search algorithm, which does not require the derivative of a function to find the minimum value of the function [43]. The algorithm uses two test points in the interval $[a, b] \subset \mathbb{R}$. The interval search becomes smaller in each episode with the end points determined by the Fibonacci sequence. The limit of the ratio of two consecutive the Fibonacci numbers converges to the golden ratio ($\frac{1+\sqrt{5}}{2}$) [44]. When the golden ratio is used instead of the Fibonacci sequence, the algorithm is called golden-section search. Algorithm 1 shows the golden-section search procedure. The Fibonacci search and golden-section search algorithm can find a local optimal point for one-dimensional function. In this paper, we used a modified version (algorithm 2) of the golden section search to minimize the two-dimensional cost function.

The inverse parabolic interpolation [45] is an optimization method that approximates the cost function with a parabola (quadratic function) and uses the minimum of the parabola as the new guess to minimize the cost function. A parabola is determined by three points. Therefore, by using the first and the last points of interval and another point, a parabola can be determined. After that, the minimum of the parabola can be computed. The minimum point of the parabola will be used instead of the point that has maximum cost among the current three points. The inverse parabolic interpolation method converges very fast if the cost function is close to a parabola. Brent combines the golden section search algorithm with the inverse parabolic interpolation method [45]. Brent's method is useful when, the cost function is close to a parabola near the optimal point and not similar to a parabola far from the optimal point. However, there is no guarantee that a passive unidirectional system behaves like a parabola near the optimal point.

**Algorithm 1** Original golden-section search
**Input:** unknown function $f(x)$, and interval of optimization $[a, b]$
**Config:** tolerance $\delta$
**Output:** optimal point $x^*$

0: Start
1: $\varphi \leftarrow \frac{1+\sqrt{5}}{2}$
2: $r \leftarrow \frac{1}{\varphi}$
3: $\Delta \leftarrow b - a$
4: while $\Delta > \delta$ then
5:    $L \leftarrow b - r\Delta$
6:    $H \leftarrow a + r\Delta$
7:    if $f(L) < f(H)$ then $b \leftarrow H$
     else $a \leftarrow L$
8:    $\Delta \leftarrow b - a$
9: return $\frac{a+b}{2}$

---

**Algorithm 2** Modified golden-section search
**Input:** unknown function $f(x, y)$, two intervals search $[x_a, x_b]$ and $[y_a, y_b]$
**Config:** tolerance $\delta_x$ and $\delta_y$
**Output:** optimal point $(x^*, y^*)$,

1: $\varphi \leftarrow \frac{1+\sqrt{5}}{2}$
2: $r \leftarrow \frac{1}{\varphi}$
3: $\Delta_x \leftarrow x_b - x_a$
4: $\Delta_y \leftarrow y_b - y_a$
5: $y' \leftarrow y_0$
6: while $(\Delta_x > \delta_x)$ or $(\Delta_y > \delta_y)$
7:    if $\Delta_x > \delta_x$ then
       $x_L \leftarrow x_b - r\Delta_x$
       $x_H \leftarrow x_a + r\Delta_x$
8:    if $f(x_L, y') < f(x_H, y')$ then
       $x_b \leftarrow x_H$
       $x' \leftarrow x_L$
   else
       $x_a \leftarrow x_L$
       $x' \leftarrow x_H$
9:    if $\Delta_y > \delta_y$ then
       $y_L \leftarrow y_b - r\Delta_y$
       $y_H \leftarrow y_a + r\Delta_y$
10:   if $f(x', y_L) < f(x', y_H)$ then
       $y_b \leftarrow y_H$
       $y' \leftarrow y_L$
   else
       $y_a \leftarrow y_L$
       $y' \leftarrow y_H$
11:   $\Delta_x \leftarrow x_b - x_a$
12:   $\Delta_y \leftarrow y_b - y_a$
13: return $(x', y')$

Another non-model based optimization approach for a dynamic problem is extremum-seeking [46-48]. The extremum-seeking approach has two components. The first component injects a slow persistently exciting signal to the closed-loop system to estimate the average of the gradient. The second component splits the fast transient response of the system dynamic and the slow steady state regulating task [49]. Therefore, the extremum-seeking closed-loop dynamics works in two timescales. The optimization is performed at the slow rate. The stability of the extremum-seeking approach for the discrete-time system is quite different from continuous time system [50]. One drawback of extremum-seeking is the transient state performance, i.e. it is slow [51]. Application of the extremum-seeking approaches is also problematic due to tuning and assignable transient performance [52]. The extremum-seeking approach has three key parameters: amplitude and frequency of the dither signal and the gain of gradient algorithm. The rate of convergence and final proximity to the optimal value are highly dependent on the parameter values. Tuning the parameters is more challenging in the presence of noises and disturbances. The assumption behind the extremum-seeking approach is that the function or system must be analytical. However, this assumption is not satisfied in the case of the unidirectional system. Therefore, we do not use the extremum-seeking approach in this paper.

Hill-climbing [53-55] is an optimization algorithm that belongs to local searches. This algorithm iteratively increments/decrements a parameter toward improving the response. The increment/decrements size is constant, which is called step size. The neighbors of each point is defined as the set of all points one step size distance along the principal axes. I.e.,

$$\mathcal{N}(x) = \{ y = x \pm \text{s}\, \vec{e}_i \,, \; i = 1,2,\ldots,n \} \tag{20}$$

where $x \in \mathbb{R}^n$ is the interested point, $y \in \mathbb{R}^n$, $\text{s} \in \mathbb{R}_{>0}$ is the step size, and $\vec{e}_i$ are mutually orthogonal unit vectors (standard basis) for the coordinate system. The neighbors set of a point has $2n$ members, where $n$ is dimension of the point. The hill-climbing algorithm starts from an initial guess (point) as the current state. In each iteration, the algorithm evaluates the current state and all point in neighbor of the current state. Then, the algorithm put the point that has maximum value, as current state. The algorithm ends when the current state is not changed, or the difference between the value of current state and the value of the maximum of points is less than a predefined threshold. In contrast to the golden section search algorithm, the rate of change in the hill-climbing algorithm is small due to the small fixed step size.

The stochastic hill-climbing algorithm (Algorithm 3) is an extension of the hill-climbing algorithm. Like the hill-climbing algorithm, the stochastic hill-climbing algorithm starts with an initial guess. Similar to the hill-climbing algorithm, in each iteration, it evaluates all points (2n points) in the neighborhood where the step size is constant. In contrast to the hill-climbing algorithm, it does not choose the maximum neighbor as next current state. It finds the direction of the maximum neighbors. Then, it generates a random number by a predefined probability density function that has a positive expected value. Next, it multiplies the random number and the direction to generate a stochastic vector. Then, it adds the current step and the stochastic vectors and use it as the current states in the next iteration. The ending of the stochastic hill-climbing algorithm is same as the hill-climbing algorithm. The stochastic hill-climbing converges to the global maximum if and only if the reward function is a convex function. Otherwise it will converge to a local maximum of the reward function. The original stochastic hill-climbing algorithm is an offline algorithm that returns the (local) optimal point and requires several function evaluations and several iterations.

**Algorithm 3** Original stochastic hill-climbing (offline)
**Input:** unknown function $f(x)$, and initial point $x_0$
**Config:** probability distribution for generating random number $\rho(v)$ and convergence threshold $\varepsilon$
**Output:** local maximum $x^*$

1: $\delta \leftarrow \infty$
2: $x \leftarrow x_0$
3: while $\delta > \varepsilon$
4: $\quad y \leftarrow x$
5: $\quad$ for all z in $NEIGHBORS(x)$
6: $\quad\quad$ if $f(z) > f(y)$ then $y \leftarrow z$
7: $\quad \delta \leftarrow f(y) - f(z)$
8: $\quad \Delta \leftarrow y - x$
9: $\quad r \leftarrow \rho(v)$ $\quad$ // generate random number
10: $\quad x \leftarrow x + r.\Delta$ $\quad$ // update x
11: $x^* \leftarrow x$
12: return $x^*$

In this paper, we do not have access to the function. Thus, we use the online version of stochastic hill-climbing that is shown in algorithm 4. In the online version, instead of finding the optimal point, the best point in the neighbors is selected for the next call. The online stochastic hill-climbing algorithm evaluate the function in the current point and all neighbors and return the maximum of them. The current point initialized by the user. Then, the return point will be used as the current points. The online stochastic hill-climbing algorithm does not have any iteration. Therefore, the algorithm should be run repeatedly whenever the quality data is available. In the case of Markovian system, the online stochastic hill-climbing will coverage similar to the offline version of the algorithm. This optimization algorithm has been used in the control system such as control of lighting [56].

**Algorithm 4** Online stochastic hill-climbing
**Input:** unknown function $f(x)$, and current point $x_{old}$
**Config:** probability distribution for generating random number $\rho(v)$
**Output:** new point in neighbor $x_{new}$

1: $y \leftarrow x_{old}$
2: for all z in $NEIGHBORS(x_{old})$ do
3: $\quad$ if $f(z) > f(y)$ then $y \leftarrow z$
4: $\Delta \leftarrow y - x_{old}$
5: $r \leftarrow \rho(v)$ $\quad$ // generate random number
6: $x_{new} \leftarrow x_{old} + r\,\Delta$ $\quad$ // update x
7: return $x_{new}$ 6: end for
7: $\Delta \leftarrow y - x_{old}$
8: $r \leftarrow \rho(v)$ $\quad$ // generate random number
9: $x_{new} \leftarrow x_{old} + r.\Delta$ $\quad$ // update x
10: return $x_{new}$

## 4. Proposed method

In the last section, the online version of stochastic hill climbing has been explained. This algorithm can be used to solve either Eq. (17) or Eq. (19). The algorithm is model-free and can be applied to any system. The algorithm evaluates the cost function at the current point as well as 4 neighbor points to select the best one. The drawback of this approach is that it requires 5 evaluations for each update. In other words, the convergence is slow.

In this paper, we propose an associated search element (ASE) to eliminate the evaluation steps. The ASE is model-based and adds expert knowledge to the optimization problem (Algorithm 5). Therefore, the proposed method is applicable to a first order discrete-time passive unidirectional system. However, convergence speed increases significantly. In the proposed method, the ASE checks the update eligibility and determines step size and direction of stochastic hill-climbing.

---

**Algorithm 5** Online stochastic hill-climbing
**Input:** parameter Y, direction of hill $D$, step size $S$, and eligibility $\mathcal{E}$.
**Config:** probability distribution for generating random number
**Output:** updated parameter Y

1: If $\mathcal{E} = 0$ then End
2: $\omega \leftarrow$ generate random number
3: $\Delta \leftarrow \omega\, S$
4: If direction is positive (upward) then
$\quad\quad Y \leftarrow Y + \Delta$
  else
$\quad\quad Y \leftarrow Y - \Delta$
5: End

---

**Algorithm 6** associated search elements
**Input:** desired trajectory buffer R, error buffer E, Controller parameters K and N
**Config:** step sizes $\{\beta_1, \beta_2, \beta_3, \beta_4, \beta_5\}$, error threshold $\epsilon$, and constant $\xi$
**Output:** direction of hill $\{D_K, D_N\}$, step sizes $\{S_K, S_N\}$, and eligibility $\{\mathcal{E}_K, \mathcal{E}_N\}$

1: Initialize $D_K \leftarrow 0, D_N \leftarrow 0, S_K \leftarrow 0, S_N \leftarrow 0, \mathcal{E}_K \leftarrow 0$, and $\mathcal{E}_N \leftarrow 0$
2: If buffers are not full then End
3: If any element of buffer R is negative then
$\quad\quad$ Dequeue buffers
$\quad\quad$ End
4: If buffer R is not ascending then
$\quad\quad$ Dequeue buffers
$\quad\quad$ End
5: If first half of buffer R is zero then
$\quad\quad$ Dequeue buffers
$\quad\quad$ End
6: $U \leftarrow K\,E + (N - K)\,R$
7: If first element of U is not positive then
$\quad\quad$ Dequeue buffers
$\quad\quad$ End
8: If all element of buffer E are negative then
$\quad\quad \mathcal{E}_N \leftarrow 1,\ D_N \leftarrow -1,\ S_N \leftarrow \beta_1$
$\quad\quad$ Jump to step 16
9: If first or second element of buffer E are not positive then
$\quad\quad$ Dequeue buffers
$\quad\quad$ End
10: If any element of buffer E is negative then
$\quad\quad \mathcal{E}_K \leftarrow 1,\ D_K \leftarrow -1,\ S_K \leftarrow \beta_2$
$\quad\quad$ Jump to step 16
11: If all element of buffer E are less $\epsilon$ than
$\quad\quad$ Dequeue buffers
$\quad\quad$ End
12: If $\max(U) > \gamma\, \text{mean}(U)$ than
$\quad\quad \mathcal{E}_K \leftarrow 1,\ D_K \leftarrow -1,\ S_K \leftarrow \beta_3$

       Jump to step 16
13: $\Omega \leftarrow -\frac{\text{mean(diff}(E))}{\text{mean}(E)}$
14: $\xi \leftarrow$ generate a random number from uniform distribution
15: if $\Omega > \rho \xi$ than
       $\mathcal{E}_K \leftarrow 1, D_K \leftarrow +1, S_K \leftarrow \beta_4$
   else
       $\mathcal{E}_N \leftarrow 1, D_N \leftarrow +1, S_N \leftarrow \beta_5$
16: Flush buffers
17: End

In the adaptive control literature, the updating task is usually applied in each time step, and the parameters are updated regardless of the quality of the data. In other words, they assume that the data are always meaningful and add information about the system. This assumption is not correct in the case of a unidirectional system. For example, a user may define a trajectory that becomes negative at some time. In this example, the unidirectional system cannot follow the trajectory. Therefore, updating parameters is meaningless and even may cause divergence. Another example is when the desired trajectory decreases faster than the dissipative rate of the unidirectional system. In this example, the error is negative. Because the regulator cannot generate negative values to a unidirectional system, updating may cause divergence. Therefore, a proper associated search element must check the quality of the data to produce eligibility signals. In addition to the data quality, the stochastic hill-climbing algorithm can update only one parameter in an iteration. Thus, only one parameter must be eligible to update ($\mathcal{E}$ is the eligible parameter in Algorithm 5).

When the algorithm finds that an updating task is not eligible for any reason, the simplest solution is to ignore the set of data and wait for the next set of data. The main drawback of this solution is that it makes the convergence slower. The better solution is to keep the last elements of the data and check if these elements and future data will be qualified for updating task (eligibility). This procedure can be implemented by buffers (queue) that act as first input first output (FIFO) memory. When the first portion of the data is not proper, the element can dequeue the data and fill the buffer with new data.

The solution of Eq. (17) depends on the set of trajectories. For example, if the set of trajectories cover only low values, the parameter $K$ increases more than optimal value $K^*$. However, if the set of trajectories cover all values (low and high), the parameter $K$ converges to optimal value $K^*$. The solution of Eq. (19) converge to the smaller values than $(N^*, K^*)$ because the regulator output has relation to the cost function. Therefore, the solution of Eq. (19) causes steady state errors. Thus, the proposed associated search element solves the following optimization problem instead of Eqs. (17) and (19).

$$\begin{bmatrix} N^* \\ K^* \end{bmatrix} = \begin{cases} \underset{N,K}{\text{argmin}} \underset{k=1,\dots,T}{\max} e(k) \\ s.t. \quad \frac{\underset{k=1,\dots,T}{\max} u(k)}{\sum_{k=1}^{T} |u(k)|} < \frac{\gamma}{T} \end{cases} \quad (21)$$

where $\gamma \in \mathbb{R}_{>0}$ is a scalar. With respect to Eq. (17), this relaxation causes that the result to converge to a point that has a higher mean error and a lower controller effort. With respect to Eq. (19), this relaxation causes the result to converge to a point that has a lower mean error and a higher mean controller effort.

Algorithm 6 shows the proposed associated search element. This element has two buffers (queue) that act as the first input first output (FIFO) memories. The buffers store regulating errors and desired states. These buffers are used to guarantee that update tasks occur only in the input direction of the system. In this algorithm, the $\beta_1, \beta_2, \beta_3, \beta_4,$ and $\beta_5 \in \mathbb{R}_{>0}$ are step sizes defined by the user i.e. hyperparameters). The five step sizes can be selected equally by the user. However, having several step sizes increases the speed of convergence. The $\rho \in \mathbb{R}_{>0}$ is a parameter that controls the probability of updating $K$ over $N$. The $\varepsilon \in \mathbb{R}_{>0}$ is the threshold for the error. The associate search element sends positive direction to the stochastic hill-climbing algorithm to increase the parameters $N$ and $K$ if and only if all the data in the error buffer is greater than the positive threshold and the ratio of the maximum of regulator output buffer to average of regulator output buffer is less than $\gamma \in \mathbb{R}_{>0}$. Therefore, the parameters are bounded. If the parameters $N$ or $K$ increase more, either the error will be negative or the ratio is more than $\gamma$. Then, the parameters decrease. Therefore, the parameters converge to a local optima. In the associated search element, we used variable $\Omega$

$$\Omega := -\frac{\max_i(\Delta M_e)}{E[M_e]}. \tag{22}$$

The definition of operator $\Delta$ is $\Delta z(n) = z(n) - z(n-1)$, where $z(n)$ is the $n$'th element of ordinal set $z$. The E[.] is expected value operator on a vector. The $M_e$ is a buffer that stores the error of the closed-loop system. The $\Omega$ converges to zero as the system approaches steady state. The stochastic hill-climbing updates only one parameter in each iteration. The proposed method uses the $\Omega$ as priority to select between the parameters. The associated search element compares $\Omega$ with a random number $\zeta$ that is distributed uniformly in the range 0 to 1. The positive scalar $\rho$ is used as a weight factor for $\zeta$. The buffers of associated search element are flushed after calling hill-climbing algorithm.

## 5. Stability

Consider the closed-loop system Eq. (12), the system is stable when error is negative $e(k) \leq 0$ because system is passive and unidirectional. In this section, we prove the stability of the proposed method when $e(k) > 0$. If the parameters $N$ and $K$ of the closed-loop system Eq. (12) are constant (non-adaptive regulator), the system is sable if and only if

$$\begin{cases} |a - bK| < 1 \\ |N| < \infty \end{cases}. \tag{23}$$

In the proposed method, we kept the parameters in the following area, which is stable according to Eq. (23)

$$\begin{cases} 0 < K < \frac{1+a}{b} \\ 0 < K \leq N < \infty \end{cases}. \tag{24}$$

Assume that $K > \frac{1+a}{b}$. If $r(k+1) + [bK - bN - a]r(k) > [bK - a]e(k)$, then $e(k+1) > 0$, which means that the system remains temporary stable. Otherwise, $e(k+1) < 0$ and parameter $K$ decreased by ASE (algorithm 6, step 9). After several decreases the system will be stable $< \frac{1+a}{b}$.

Another unstable condition happens when $N > K + \frac{1-a}{b}$ and $N$ is increasing. From algorithm 6, we can conclude that $N$ is increasing only when $e(k) > 0$ for the all of last $\mathcal{T}$ steps. When $e(k) > 0$, from (12) we have

$$e(k+1) = [a - bk]\,e(k) + r(k+1) + [bK - bN - a]r(k). \tag{25}$$

We can show $N > K + \frac{1-a}{b}$ by

$$N = K + \frac{1-a}{b} + \Delta \tag{26}$$

where $\Delta \in \mathbb{R}_{>0}$. By substituiting Eq. (26) in Eq. (25)

$$e(k+1) = [a - bk]\,e(k) + r(k+1) - \left[bK - b[K + \frac{1-a}{b} + \Delta] - a\right] r(k). \tag{27}$$

Therefore,

$$e(k+1) = [a - bk]\,e(k) + r(k+1) - [1 + b\,\Delta]\,r(k). \tag{28}$$

Let us define $\check{a} := a - bk$ and $\check{b} := 1 + b\,\Delta$. Thus, $|\check{a}| < 1$, and $\check{b} > 1$. By this definition, (28) can be written as

$$e(k+1) = \check{a}\,e(k) + r(k+1) - \check{b}\,r(k). \tag{29}$$

By defining $m(k) := -r(k+1) + \check{b}\,r(k)$,

$$e(k+1) = \check{a}\,e(k) - m(k). \tag{30}$$

When the desired trajectory is bounded, $r(k+1) \leq r_{max}$. Therefore,

$$m(k) \geq -r_{max} + \check{b}\, r(k) \,. \tag{31}$$

We have established that $K$ is bounded, therefore

$$\lim_{N \to \infty} m(k) = \lim_{\Delta \to \infty} m(k) = \lim_{\check{b} \to \infty} m(k) = +\infty. \tag{32}$$

Therefore, there is an upper limit $N_{max}$ such that

$$\forall N > N_{max} \implies m(k) \geq 0. \tag{33}$$

Also, according to Algorithm 6, $N$ increases if buffer $R$ is ascending. Thus, we can conclude

$$\forall N > N_{max} \implies m(k) \geq M > 0. \tag{34}$$

By substituting Eq. (34) in Eq. (30)

$$e(k+1) \geq \check{a}\, e(k) - M \,. \tag{35}$$

From Eq. (35), we can conclude that

$$\exists \check{k} \; s.t. \; \forall k > \check{k} \implies e(k) < 0. \tag{36}$$

Equation (38) contradicts the condition of increasing parameter $N$ in Algorithm 6, i.e. $e(k) > 0$. Thus Algorithm 6 part 7 decreases the parameter $N$.

The third unstable condition happens when $K < K^* < \frac{a}{b}$ and $K$ is decreasing toward zero. When parameter K is very small, $a - bK \approx a$. Then, we can rewrite (12) as

$$e(k+1) = \begin{cases} a\, e(k) + r(k+1) + [-bN - a]\, r(k) &, e(k) > 0 \\ f\, e(k) + r(k+1) - fr(k) &, e(k) \leq 0 \end{cases}. \tag{37}$$

From algorithm 6, we can conclude that $K$ is decreasing when $e(k) > 0$ in first two steps and then $e(k) < 0$ for some steps in the last $\mathcal{T}$ steps. The sign of $e(k)$ changes when $a\, e(k) + r(k+1) < [bN + a]\, r(k)$. As a results

$$r(k) > \frac{a}{bN+a} e(k) + \frac{1}{bN+a} r(k+1). \tag{38}$$

From Eq. (38) we can conclude that $r(k) > r(k+1)$, which has contradiction with algorithm 6, step 3. Thus, this condition never happen.

## 6. Simulations

To evaluate the proposed method, consider a TCP actuator, as an example of unidirectional first order system, with $a = f = 0.98195$, $b = 0.00042345$ and the time step $T = 0.1$ second. These practical magnitudes are taken from prior literature on TCP actuators [57]. All simulations were implemented in Python 3.6.4. The three methods presented are stochastic hill climbing (SHC), golden section search (GSS), and the proposed method (SHC with ASE). A uniformly distributed random number generator is used for all the simulation.

Fig. 2 shows four simulation results in force regulating of the TCP actuator when the trajectory is periodic. Hyperparameters in the proposed method are selected as $\beta_1 = 2, \beta_2 = 2, \beta_3 = 2, \beta_4 = 5, \beta_5 = 10$, $\gamma = 1.1$, and $\varepsilon = 0.01$. Fig. 2 (a) shows an example of this system when the parameters, $N$ and $K$, are initialized zero, and the trajectory is chosen as a rectangular pulse $r(k) = Heaviside\left(\sin\left(\frac{2\pi T}{\tau} k\right)\right)$, (100 s period and duty cycle 50%) such that the response of the system converges to zero when the input is zero. The system with SCH and the proposed method could not reach the desired trajectory at the beginning (initial steps). After some updates, the proposed method gets closer to the trajectory. The golden section search (GSS) converges faster than the other algorithms. The SHC regulator and the proposed method generate control output ($u$) to improve the system response; however, the regulator output remains bounded. The regulator output of the GSS becomes very large at the beginning because K and N take high values, which may cause a damage to the physical system. The parameters $N$ and $K$ increase in the SHC and the proposed method, but it is not as significant as the magnitudes in the case of GSS. The proposed method converged to the local optima in about $4^{th}$ episodes; however, the SHC did not converge even in the $6^{th}$ episode. The GSS search is faster than the other methods in convergence (only one episode). Fig. 2 (b) demonstrates a simulation similar to the previous simulation, except that the initial values are selected large ($K = 450$ and $N = 600$). At the beginning, there is some amount of overshoot in each method. All the methods successfully decreased the parameters ($N$ and $K$) until they reach to the desired values. The GSS converges in one episode, and the other methods converge in the $5^{th}$ episode. Fig. 2 (c) illustrates a simulation when the trajectory is a versine wave $r(k) = \text{versine}\left(\frac{2\pi T}{\tau} k\right) = 1 - \cos\left(\frac{2\pi T}{\tau} k\right)$, with period of 200 s and parameter N and K started from zero. The proposed method converges in the $4^{th}$ episode. The golden section search converges in the first episode and SHC does not converge. As Fig 2 (d) shows, when the parameters are initial large ($K = 300$ and $N = 500$), the proposed method also converges.

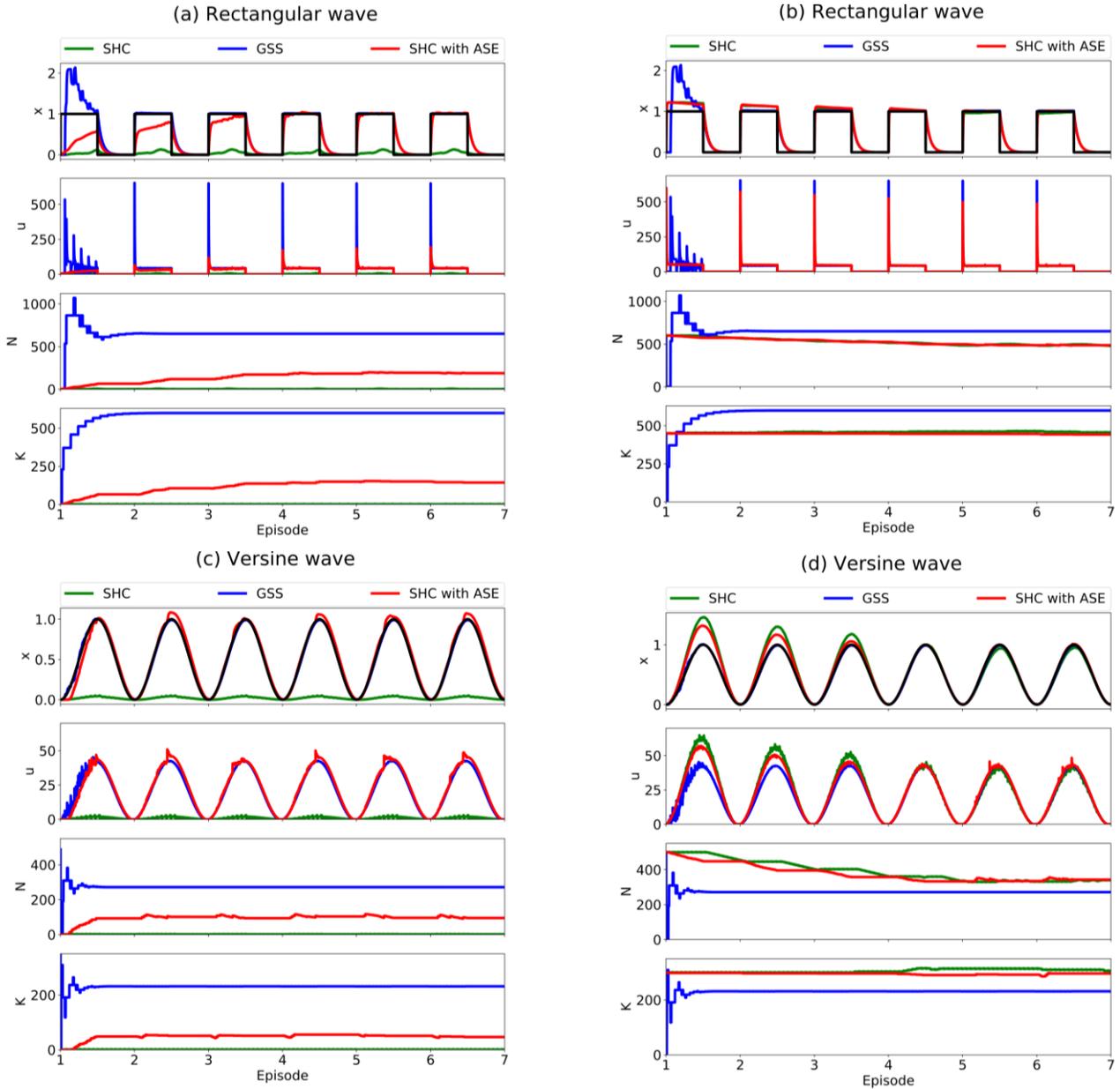

Fig. 2. Simulation of proposed method with periodic trajectories (a) rectangular wave when the parameters, $N$ and $K$, are initialized zero (b) rectangular wave when the initial values of the parameters are selected large, $K = 450$ and $N = 600$ (c) versine wave when the parameters, $N$ and $K$, are initialized zero (d) versine wave when the initial values of the parameters are selected large, $K = 300$ and $N = 500$.

Fig. 3 in shows that the proposed method not only works with a periodic trajectory but also successfully works in other conditions. The hyperparameters in the proposed method are the same as before ($\beta_1 = 2, \beta_2 = 2, \beta_3 = 2, \beta_4 = 5, \beta_5 = 10, \gamma = 1.1$, and $\varepsilon = 0.01$). The trajectory in Fig. 3 (a) and (b) are a random step $r(k) = \sum_{n=0}^{\infty} \rho_n [Heaviside\left(k - \frac{n\tau}{T}\right) - Heaviside\left(k - \frac{[n+1]\tau}{T}\right)]$, in which each step takes 50 s. Parameters N and K are initialized as zero in for the trial in Fig 3 (a) and selected large ($K = 350$ and $N = 500$) for the trial in Fig 3 (b). The trajectory in Fig. 3 (c) and (d) are a versine with random amplitude $r(k) = \sum_{n=0}^{\infty} \rho_n versine\left(\frac{2\pi T}{\tau} k\right) [Heaviside\left(k - \frac{n\tau}{T}\right) - Heaviside\left(k - \frac{[n+1]\tau}{T}\right)]$ and each episode takes 200 s. Parameters $N$ and $K$ initialized zero in Fig 3 (c) and selected large ($K = 200$ and $N = 400$) for Fig 3 (d). GSS converges in the 2$^{nd}$ episode in random steps trajectory and converge in one episode in random versine trajectory. The proposed method converges in the 3$^{rd}$, 5$^{th}$, 2$^{nd}$, 4$^{th}$ episode respectively in Fig. (a), (b), (c), and (d). The GSS converges faster than proposed method in all the cases. The SHC fails because it updates the parameters regardless of the quality of the data. It can be concluded that the SHC cannot be used without an associated search element that generates eligible signals. Therefore, we proposed we proposed SHC with SCH to improve the system response.

Fig. 4 illustrates two simulations when the trajectory is arbitrary. The parameter $N$ is initialized 600. The parameter $K$ is initialized 460 and 470 in Fig. 4 (a) and (b). The proposed method and the SHC converge to a proper local minimum. However, the golden section search converges to ill local minima. This figure shows that that the GSS is very sensitive to initialization. It can be concluded that when the parameters reach the local optima, the proposed method can handle any arbitrary trajectory in the controllable direction. However, in the other direction (returning direction in the case of TCP muscles or SMA actuators), the trajectory should be slow such that the uncontrollable direction of the unidirectional system can follow it.

Fig. 5 shows the behavior of the methods when we have a have a prior estimate of the system parameters (i.e, $a$ and $b$) with uncertainty and when we do not know about the system but we assume an upper bound for $N$ and $K$. If we know that $a \in [a_{min}, a_{max}]$ and $b \in [b_{min}, b_{max}]$, we can infer $K \in [K_{min}, K_{max}]$ and $N \in [N_{min}, N_{max}]$. As the uncertainty decreased, the intervals decreased, and all methods converges faster. If uncertainty is low (for example 5%), we do not need any adaptive regulator. Fig. 5 (a) and (b) shows the GSS and proposed method when we do not have the estimate. Fig. 5 (c) and (d) show the GSS and proposed method when we have 20% uncertainty. The GSS has less overshoot when we have the prior estimate. There remain some steady state error in GSS as it shown in Fig 5 (c). In Fig. 5 (d), the proposed method keeps the parameters about the initial value because the 20% uncertainty is enough to regulate the system with zero error.

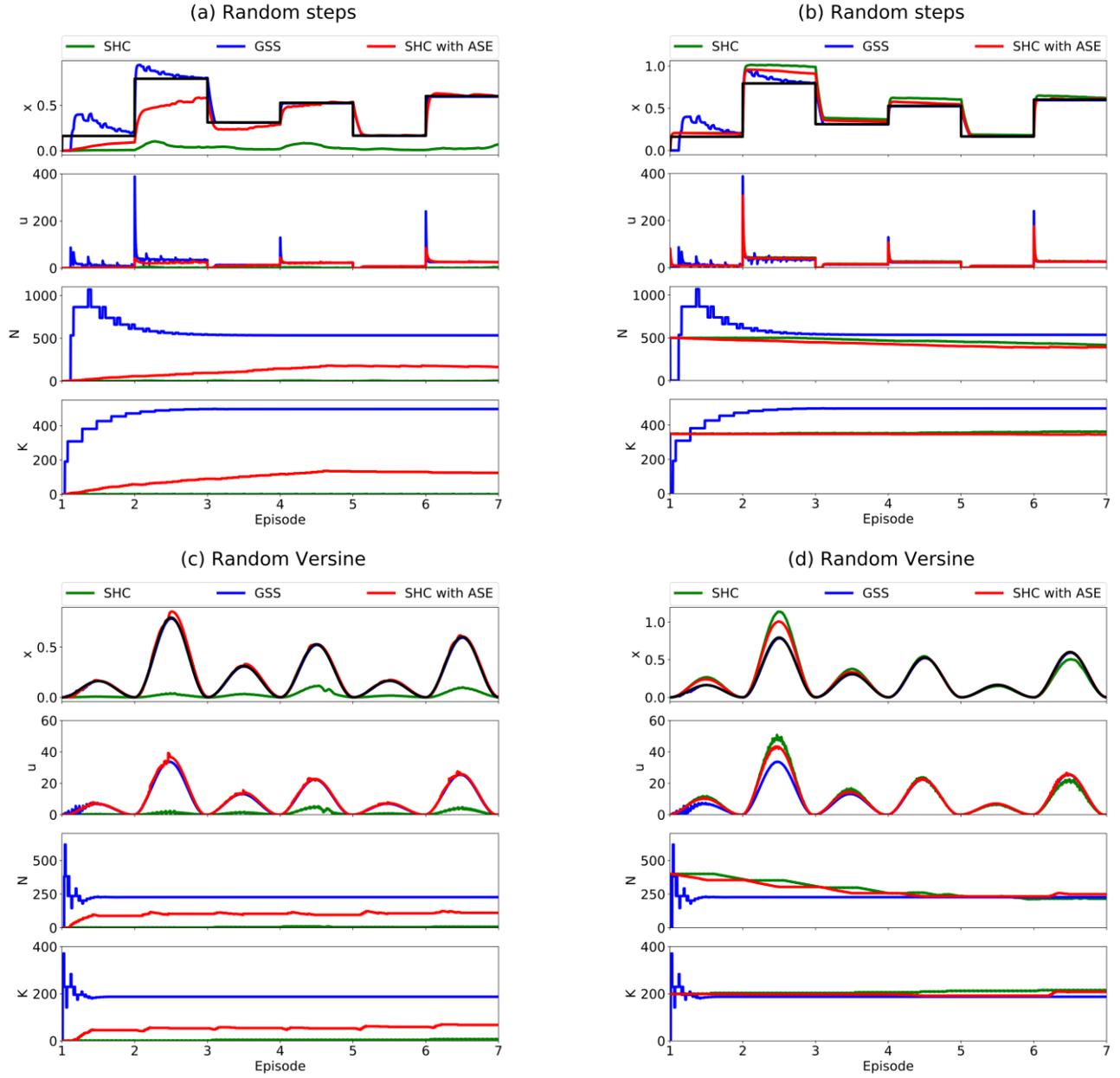

Fig. 3. Simulation of proposed method with non-periodic trajectories (a) random steps when the parameters, $N$ and $K$, are initialized zero (b) random steps when the initial values of the parameters are selected large, $K = 350$ and $N = 500$ (c) random versine when the parameters, $N$ and $K$, are initialized zero (d) random versine when the initial values of the parameters are selected large, $K = 200$ and $N = 400$.

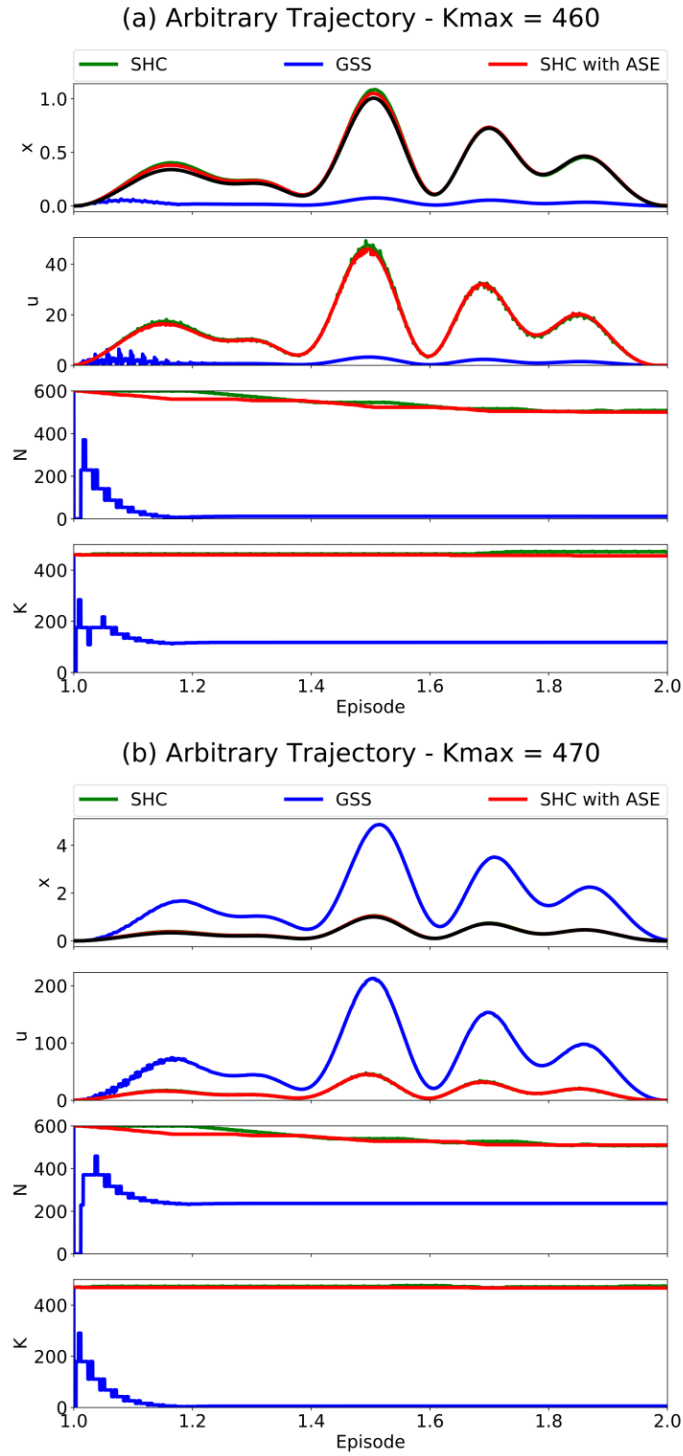

Fig. 4. Simulation of proposed method with arbitrary trajectory the parameter are initialized as $N = 600$ and (a) $K = 460$ (b) $K = 470$.

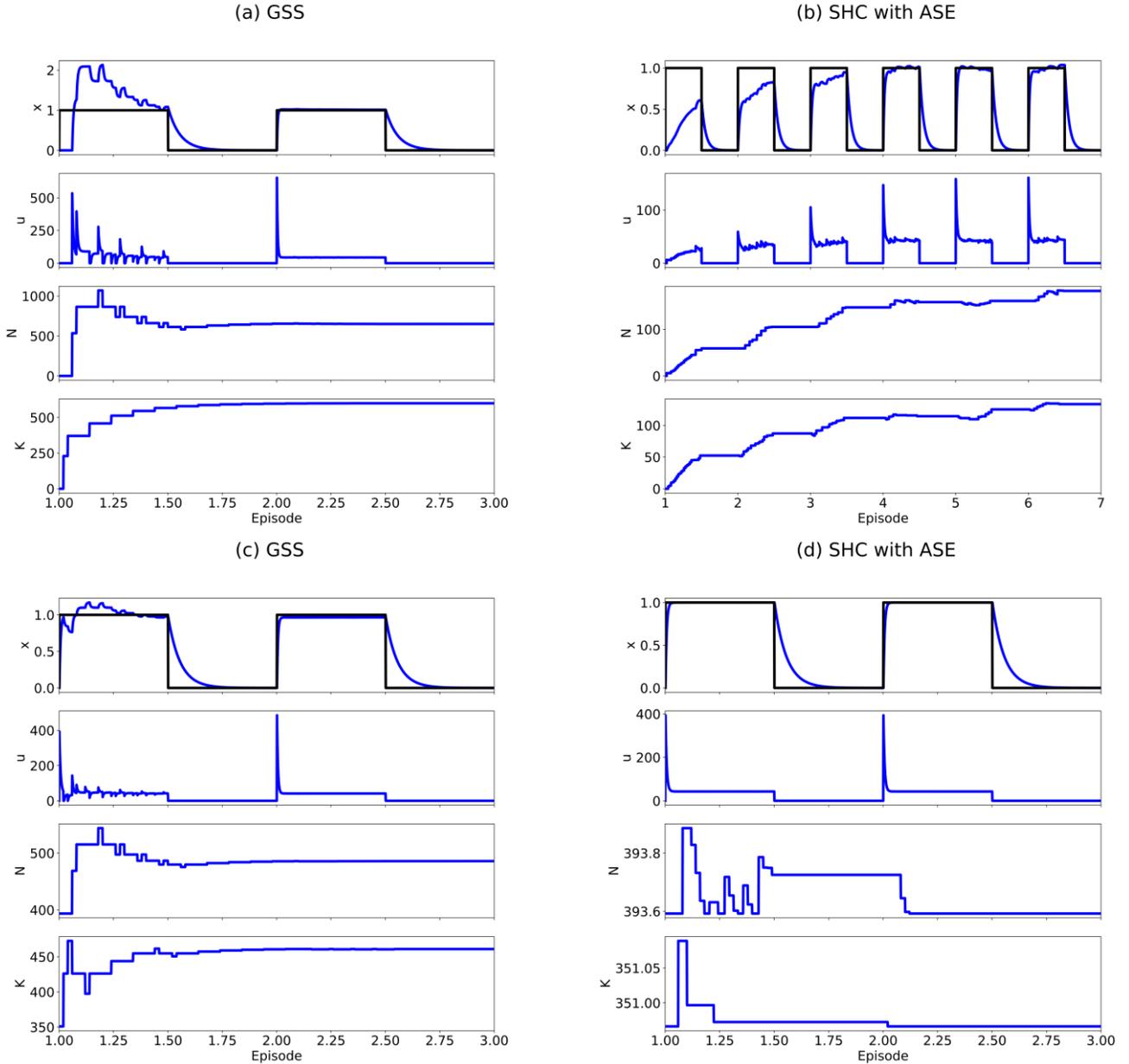

Fig. 5. Effect of having an estimation of system parameters $a$ and $b$ (a) simulation of golden section search without having an estimation and parameter initialized as zero (b) simulation of proposed method without having an estimation and parameter initialized as zero (c) simulation of golden section search with having an estimation of system parameter with 20% uncertainty such that we can have a better initial values of $K$ and $N$ (d) simulation of proposed with having an estimation with 20% uncertainty.

## 7. Experimental Results

Our method works for any first-order passive unidirectional system, as discussed before. After demonstrating the ability of proposed method in the simulation, we have done several experiments with a TCP muscle and an SMA muscle. Fig. 6 demonstrates the experimental setup, which includes an NVIDIA Jetson TX2 Developer kit, a power supply (Topward 6306D), a Minibeam load cell (Omega LCEB-5), an instrumentation amplifier (Texas Instrument INA 122) with gain of 2000, an analog to digital converter (TI ADS1115), a digital to analog converter (Microchip Technology MCP4725), and a power amplifier circuit (Texas Instrument OPA2244 and ST Microelectronics TIP122) with voltage gains of 4.09. The load cell has 3 mv/v rated output, 7 mN linearity, 5 mN hysteresis, and 3 mN repeatability. All regulators were implemented with Python 3 on the NVIDIA Jetson TX2 developer kit. The sampling time was 0.1s.

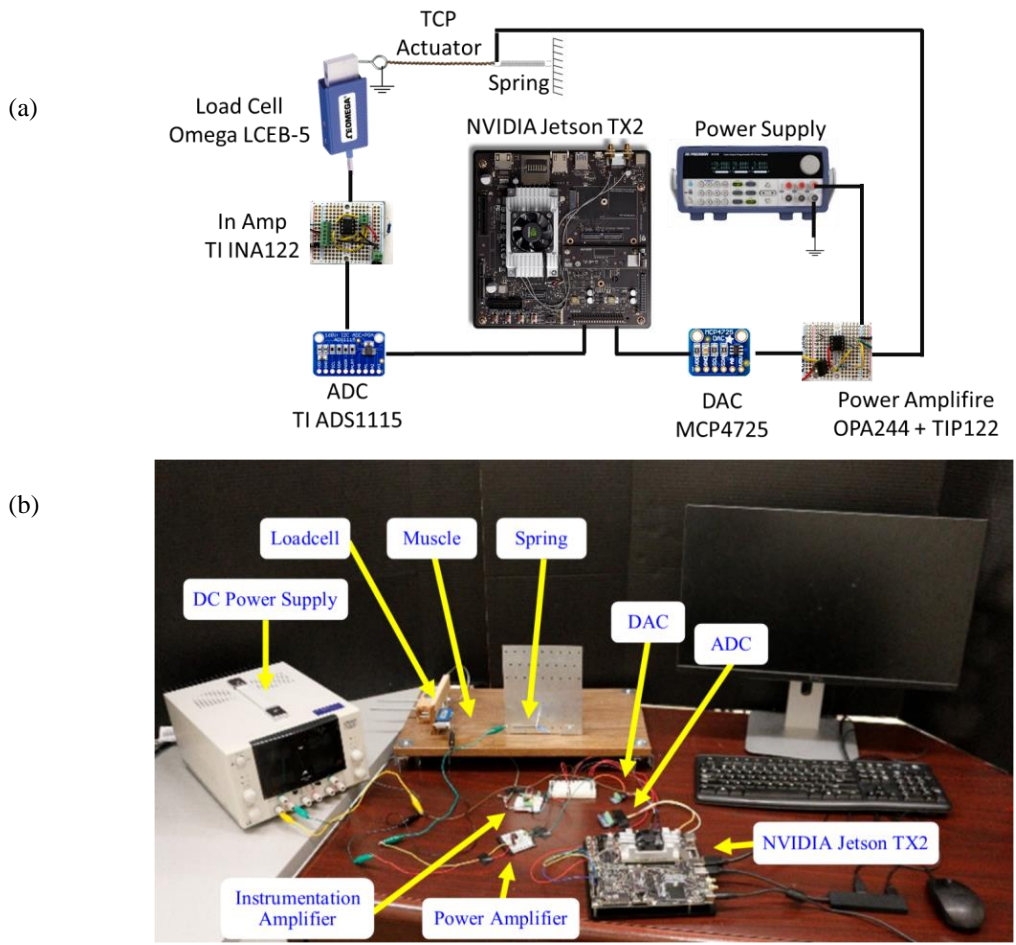

Fig. 6. Experimental setup for force regulating of an 1-ply TCP actuator or SMA muscle (a) schematic (b) photo.

Adaptive controllers can improve performance in the case that the system dynamics change over the time. In our previous publications, we showed that a first-order, discrete-time, state-space model is sufficient for regulating TCP muscles [22, 57]. Fig. 7 shows the response of a 1-ply TCP muscle when the input is a rectangular wave with a period of 120 s and duty cycles of 50%. We use this trajectory because, in this timing, the TCP muscle reaches steady-state, which is helpful for system identification processes. The TCP muscle was 10 cm and pre-stretched with 1 N, using a spring with stiffness 52.538 N/m. In this experiment, we used a 12V Lithium Iron Phosphate (LiFePO4, LFE) battery. We paused the experiment and recharged the battery two times. In the first battery cycle, the actuator worked for 884.7 minutes, as shown in Fig. 7 (a). Next, the battery was changed (as the battery charge indicator was low) and the actuation experiment was performed. The result is shown in Fig. 7 (b) from 884.7 to 1995.6 minutes. In the third battery cycle, the muscle broke at 2199.8 minutes, as shown in Fig. 7(c). The measured forces for each battery cycle are shown in Fig. 7 (a), (b), and (c), in order. In Fig 7. (c), the muscle broke, and we stopped the recording. As shown in Fig. 7 (a), the behavior of muscle changed slowly over time, which is the main observation from the test result. We selected the first, the middle, and the last 20 cycles of the first battery cycle, Fig 7. (a), and plotted it in Fig. 7 (d), (e), and (f), in order. Next, we performed system identification using MATLAB toolbox by selecting canonical observable discrete-time state-space with sampling time 0.1s and prediction error method (PEM). The simulated results of these cycles are presented in Fig. 7 (g), (h), and (i) in order. The results show that the parameters of a, b, and DC gain changes slowly over time. Therefore, we should repeat system identification numerous times, or we design an adaptive regulator. The regulator, $r(k+1) \approx r(k)$, is useful for many cases, such as tracking the sun in solar panels or pick and place application with gripper/hand.

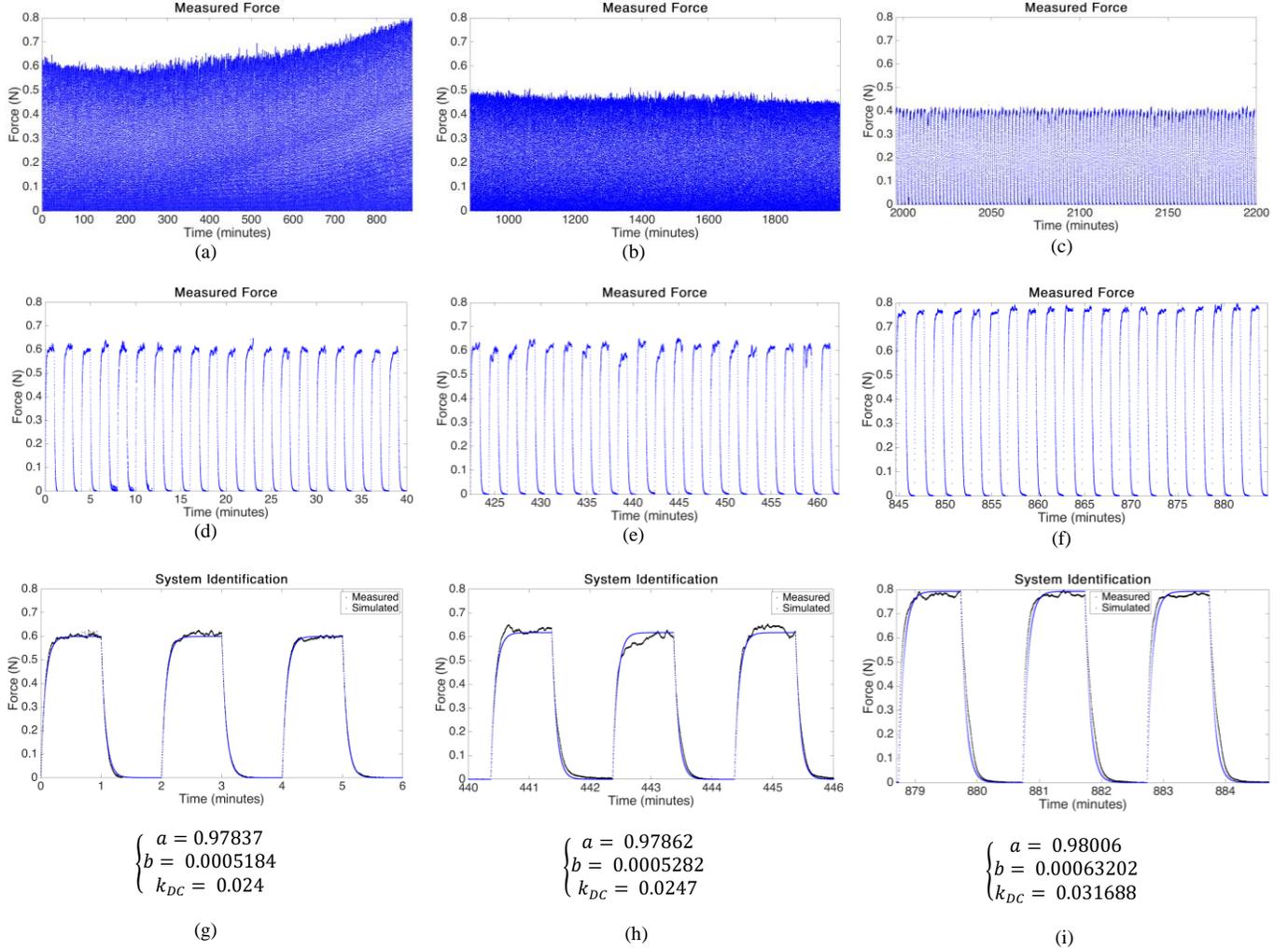

(g)
$\begin{cases} a = 0.97837 \\ b = 0.0005184 \\ k_{DC} = 0.024 \end{cases}$

(h)
$\begin{cases} a = 0.97862 \\ b = 0.0005282 \\ k_{DC} = 0.0247 \end{cases}$

(i)
$\begin{cases} a = 0.98006 \\ b = 0.00063202 \\ k_{DC} = 0.031688 \end{cases}$

Fig 7. Response of a 1-ply TCP when input is rectangular wave with period of 2 minutes and duty cycle of 0.5 (a) the first experiment (b) the second experiment (c) the last experiment until the muscle brakes (d) measured force in the first 20 cycles of the first experiment (e) measured force in the middle 20 cycles of first experiment (f) measured force in the last 20 cycle of first experiments (g) system identification of the first 20 cycles of the first experiment (h) system identification of the middle 20 cycles of the first experiment (i) system identification of the last 20 cycles of the first experiment.

Fig. 8 shows the experimental result in regulating force of TCP actuator as an example of passive first order system. The left column shows an experiment when the trajectories are a rectangular wave. Each episode was 2 minutes ON and 2 minutes OFF, with magnitude of 125 mN. This timing is selected because we want to test the method in a condition like a step input. The initial value for parameter $N$ and $K$ was small to avoid breaking the TCP actuator. After 6 episodes, the parameter $K$ converged to 50 and parameter $N$ converged to 100. We can conclude that the proposed method converges in rectangular wave and then can be used in a step trajectory. The middle column demonstrates the experimental result when the trajectories are a versine function. In contrast to the rectangular wave, the versine function has only one frequency; however, it does not have a flat portion to provide condition similar to a steady state. The desired trajectory has the period of 1000 seconds and magnitude of 100 mN. This timing is selected because the actuator must have enough time to cool down and follow the trajectory (i.e. physical limitation of TCP actuator). The initial value for parameter $N$ and $K$ was small to avoid breaking the TCP actuator. After 6 episodes, the parameter $K$ converged to 50 and parameter $N$ converge to 100. We can conclude that the proposed method converges to versine function, where there is not any steady state like condition. The right column illustrates the experimental result when the trajectories are arbitrary. Each episode takes 60 s. Like other experiments, the initial values for parameters $N$ and $K$ were small. After 45 episodes the parameter $K$ converged to 50 and parameter $N$ converge to 90. We can conclude that the proposed method converges even with arbitrary trajectory. We could not find a published paper on adaptive control of TCP muscles. There are other classical controllers, such as PI and PID proposed, but PI and PID are not adaptive controllers. Other papers on control of TCP muscle only show results for following short-time trajectories with negligible error. However, they did not show results on long-term trajectories. Their methods may produce notable error in long-term trajectories because TCP parameters change overtime as shown in Fig. 7. The performance of the regulator after convergence stated in Table 1.

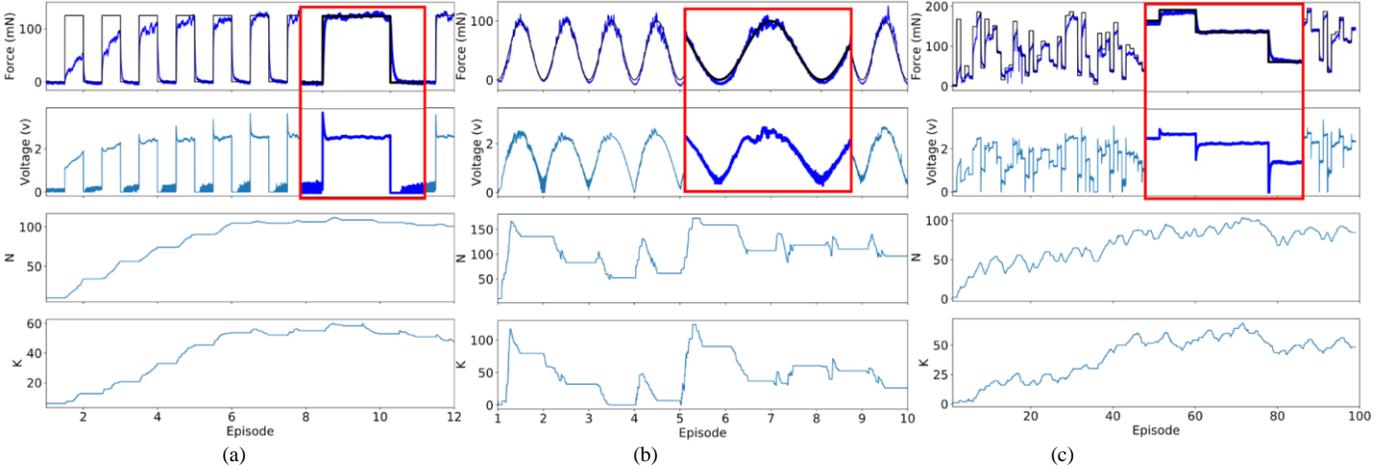

Fig. 8. Experimental result for regulating force of TCP actuator using the proposed method (SCH with ASE): (a) rectangular trajectory, (b) versine and (c) arbitrary trajectory.

| Trajectory | Rectangular | Versine | Random step |
|---|---|---|---|
| Mean absolute error (mN) | 0.0064951 | 0.0055327 | 0.0092948 |
| Root mean squared error (mN) | 2.374 | 1.1946 | 2.6165 |
| Maximum voltage | 3.5945 | 3.0941 | 3.7108 |
| Average voltage | 1.4717 | 1.677 | 1.7967 |

Table 1. Regulator performance after convergence for TCP experiments shows in Fig. 8.

To test the stability of the proposed method, several experiments on a higher order passive system with hysteresis (SMA muscle, diameter = 100 μm, obtained from Dynalloy Inc.) have been done. Fig. 9 illustrates the experimental result in regulating the force of SMA muscle. The left column shows experiment when the desired trajectory is rectangular wave. Each episode was 30 s ON and 30 s OFF with magnitude of 700 mN. The proposed method is a regulator, not tracker. If the trajectory change very fast (10s or 5s), the regulating assumption is not valid. Similar to the TCP experiment, the initial values have been elected small. After 30 episodes, the parameter *K* converged to 4 and parameter *N* converge to 8. We can conclude that the proposed method converges to a local minimum when the system has hysteresis. The middle column shows experiment when the desired trajectory was versine function with period of 60 s and magnitude of 1000 mN. Similar to TCP experiment, the initial values have been elected small. After 10 minutes the parameter *K* converged to 14 and parameter *N* converge to 18. This experiment on SMA muscle, confirm previous conclusion that the proposed method converges to a local minimum when the system has hysteresis. The right column shows the experimental result when the desired trajectory is a sequence of random steps (40 s each episode). Similar to the TCP experiment, the initial values have been selected small. The method converged to a local minima several times. However, by changing the trajectory, the method converges to a new local minimum. We can conclude that the proposed method remains stable with a system that has hysteresis, although it converges to a different local minimum. The proposed method will not converge to an optimal point in some cases of SMA actuator because it has hysteresis. The hysteresis cause infinite number of local optimum. The hysteresis does not move the optimal points when the trajectory is rectangular wave or versine. Therefore, the proposed method works (converges) in these type of the trajectory. The performance of the regulator after convergence stated in Table 2.

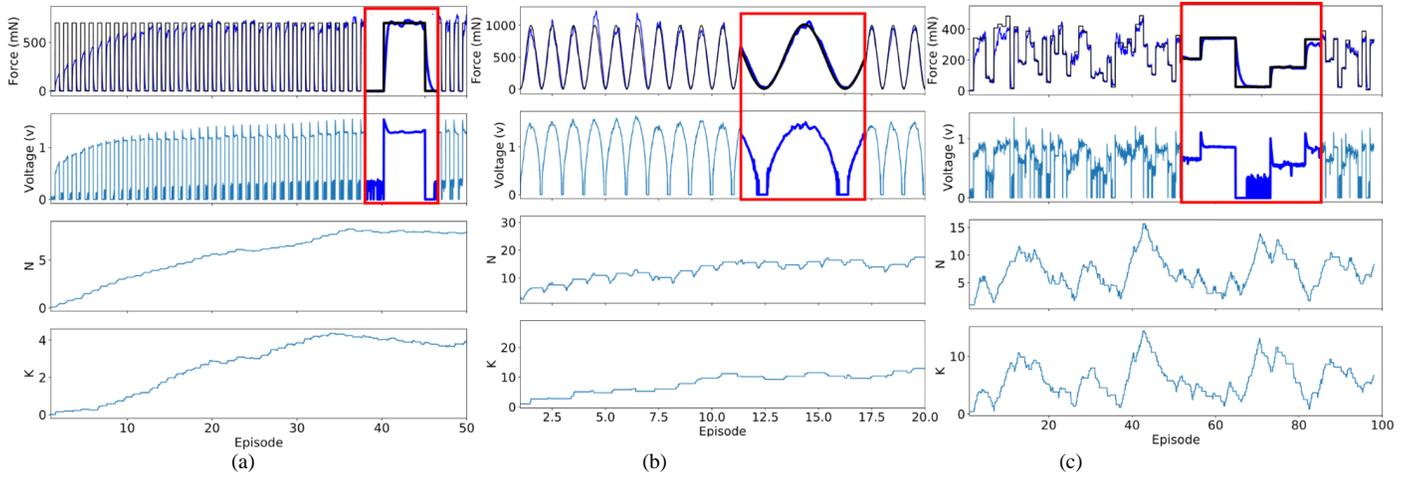

Fig. 9. Experimental result for control force of SMA muscle using the proposed method (SCH with ASE): (a) rectangular trajectory, (b) versine and (c) arbitrary trajectory.

| Trajectory | Rectangular | Versine | Random step |
|---|---|---|---|
| Mean absolute error (mN) | 0.059566 | 0.025508 | 0.024888 |
| Root mean squared error (mN) | 11.965 | 5.5683 | 4.8007 |
| Maximum voltage | 1.5676 | 1.6566 | 1.3672 |
| Average voltage | 0.79331 | 0.9207 | 0.6195 |

Table 2. Regulator performance after convergence for SMA experiments shows in Fig. 9. For random step trajectory we used the second half of data because it did not converge.

## 8. Conclusion

Unidirectional passive systems are dissipative systems with inputs that can act only in one direction and rely on dissipation to move in the other direction. In this paper, the discrete-time state space representation is used to model the system by a difference equation. Adaptive controllers can improve the system performance when the parameters of the system change over time. In this paper, a novel method is proposed to optimize the state-feedback regulator parameter online. The proposed method works for any first-order passive unidirectional system and it is general and can be applied to a wide variety of systems.

The proposed method combines expert knowledge with stochastic hill-climbing. The expert knowledge appears in the associated search element. The proposed method works only in first-order passive unidirectional systems. The stability of proposed method is proofed mathematically and demonstrated in simulations and experiments.

The comparison of three different methods was presented to illustrate the behavior in different cases. The simulation results show that the proposed method converges to local optima. In some simulation, the SHC fails because it updates the parameter when the quality of data is poor. These examples show that the associated search element is necessary in the case of passive unidirectional systems. GSS is faster than the proposed method (about 5 times faster); however, the GSS may converge to local optima such that the steady state error remains. We can conclude that to avoid the steady-state error, the proposed method (SCH&ASE) can be used.

To examine the ability of the proposed method in physical systems in presence of noise and disturbance, several experiments have been done on TCP and SMA artificial muscles. The experimental results show that ability of the proposed method in following rectangular waves, versine functions, and arbitrary trajectories.

The main drawback of the proposed method is convergence time of approximately 50 episodes. This problem should be addressed in future works. In the proposed method, the step size is fixed. If large step sizes are selected, the algorithm does not converge. If the step sizes are small, the converging time increases. In the future, we will investigate using Eq. (12) and Eq. (17) to find a new optimization method for tracking, $r(k+1) = z(k)r(k)$. Also, we will design a model predictive tracking controller when we know the $m$ future time step trajectory, $\{r(k), r(k+1), ..., r(k+m)\}$. Initial guess directly affects the convergence time. In the future, we will find a method to initialize the parameter, and then using the proposed method to fine-tuning.